\documentclass[journal,11pt,draftcls,onecolumn]{IEEEtran}
\usepackage{amsmath,amssymb,amsbsy,amsthm,geometry}
\usepackage{algorithm} 
\usepackage{algpseudocode} 
\usepackage{graphicx}
\usepackage{capt-of}
\graphicspath{{figures/}}

\let\epsilon\varepsilon
\let\phi\varphi

\newtheorem*{lemma*}{Lemma}
\newtheorem{theorem}{Theorem}

\newtheorem{claim}{Claim}

\usepackage[final,authormarkup=none]{changes}
\definechangesauthor[name={Ivan Baksheev},color=blue]{IB}

\usepackage[hidelinks]{hyperref}

\begin{document}

\title{Error-correcting codes without parity-check bits} 

\author{Boris Ryabko$^{1,2,3}$, Ivan Baksheev$^{1,3}$,
    and Vyacheslav Zhuravlev$^{1}$\\[0.5ex]
 {\small
$^{1}$Federal Research Center for Information and Computational
Technologies}\\
{\small$^{2}$Siberian State University of Telecommunications and
Informatics}\\
{\small $^{3}$Novosibirsk State University}\\
  {\small
Novosibirsk, Russia}\\
{\small
\texttt{boris@ryabko.net,
i.baksheev@g.nsu.ru, slawajur@mail.ru}}}
\date{}

\maketitle

\begin{abstract}
We are developing a new approach to the problem of error
correction in communication channels, in which the input
sequence is transformed in such a way that the interdependence
of symbols increases significantly.
Then, after the sequence is transmitted over the channel, this
property is used for error correction, so that the residual
error rate is significantly reduced.
A natural example of its application is the transmission of
binary sequences whose entropy (per symbol) is less than one,
since in this case the length of the transmitted sequence
remains constant during encoding and decoding.

An algorithm has been developed to implement the maximum
likelihood method, in which the encoding and decoding times are
linear. This algorithm was used to experimentally investigate
its effectiveness on popular channel models, and it was found
that there are certain sets of parameters for which the proposed
method is quite effective.%
\footnote{Some of the results were presented at the 2026 IEEE International Multi-Conference on Engineering, Computer and Information Sciences (SIBIRCON).}
\end{abstract}

\section{Introduction} 

Since the publication of Claude Shannon's famous article
\cite{sh}, the problem of efficient data transmission over noisy
channels has occupied a central place in information theory.
Hundreds of articles are published annually on various error
correction methods, and various monographs, such as \cite{b1,bb},
are devoted to this topic.

Many error-correcting codes studied in theory and used in
practice are based on adding redundant symbols (usually parity
bits) to the transmitted message. This class of methods includes
block codes \cite{b1,b2,wb}, convolutional codes \cite{b4}, BCH
codes \cite{tr}, low-density parity-check codes \cite{jo},
hash-function-based codes \cite{r}, and many others.

Another popular approach to transmission through a noisy
channel is a so-called joint source-channel coding (JSCC) (see
for review \cite{ve,gu}). In particular, this approach considers
codes in which data redundancy is used to correct errors in
noisy channels.

It is worth noting that when it is necessary to transmit
low-entropy data over a noisy channel, it seems quite natural to
first compress the message and then apply error-correcting
coding. In the next section, we will outline this approach and
show that problems related to loss of synchronization may arise
when decoding the received sequence. Thus, it turns out that in
a number of situations, it is better to omit the data
compression step and look for other methods of joint
source-channel coding.

In this article, we propose the following method of joint
source-channel coding. The transmitted sequence is transformed
in such a way that the interdependence of the symbols increases
significantly, and the transformed sequences themselves follow a
uniform distribution.
Then, after the sequence is transmitted over the channel, these
properties are used for error correction, which allows for a
significant reduction in the residual error rate.
Of course, this is possible only if the entropy of the
transmitted binary sequences (per symbol) is less than one---that
is, if the source has redundancy.

It is important to note that the encoding and decoding times in
the proposed method are linear (i.e., the rate is constant);
however, the required memory capacity can be quite large. So,
taking into account that the encoding and decoding rates remain
constant, it can be said that the method allows for a reduction
in the error rate ``at no cost.''

The transformation of the sequence is based on so-called
two-faced processes (see \cite{rm} for definitions).
After the transformed sequence is transmitted over the channel,
the maximum likelihood method is used to estimate the
transmitted sequence. An algorithm has been developed to
implement the maximum likelihood method, in which the decoding
time is linear. This algorithm was used to experimentally
investigate its effectiveness on popular channel models, and it
was found that there are certain sets of parameters for which
the proposed method is quite effective.

The rest of the paper covers the following. The next section contains preliminary information and background details on ``two-faced processes.'' The subsequent sections describe the encoder and decoder, respectively. This is followed by the experimental results and a discussion of possible generalizations. The concluding section presents a brief summary.

\section{Preliminaries}

\subsection{Some error-correcting codes for low-entropy sources}
As Shannon noted in \cite{sh}, it is quite natural to
use the redundancy in transmitted data to correct errors that
occur when the data is transmitted over a channel subject to
noise. As we mentioned earlier, there are a number of studies in
which this natural idea has been applied to various
``source--channel'' models.
Among these works, it is worth noting \cite{jj}, in which the
author examines Markov sources with high redundancy and a
discrete memoryless channel. He describes a method in which the
source's redundancy is used to combat errors in the channel. The
idea of using data redundancy to correct channel errors was
applied by a number of other authors in the 1990s \cite{j1,j5}
and is currently being developed within the JSCC.

\subsection{A schematic overview of the ``compression first, then error correction'' method}

The proposed method can be applied when the entropy of a binary
information source is less than one (per symbol). Therefore, we
will briefly mention other options for transmitting data with
low entropy.

At first glance, it seems natural to first use a traditional
lossless data compression method and then apply an error-correcting
code. The receiver first performs error-correcting decoding and
then ``unpacks'' the resulting sequence. However, this raises a
``hidden'' problem known as ``error propagation'' (or ``loss of
synchronization,'' or ``synchronization catastrophe'').
The essence of the problem is this: when using lossless data
compression methods, it is necessary to use codewords of varying
lengths (otherwise, lossless compression is impossible!).
However, using codewords of different lengths can lead to a
loss of synchronization: in other words, even a single error
remaining after the error correction stage can trigger a whole
cascade of errors during decompression due to the loss of
synchronization.

Consider, for example, an information source that generates the
sequence $x = x_1 \dots x_n$ ($n \ge 1$). Suppose we use a noisy
channel and an error-correcting code with a bit error rate (BER)
$= \alpha > 0$ (i.e., when this code is applied to $x$, the
average number of corrupted symbols is $n\alpha$). If the
original sequence was losslessly compressed into $y = y_1 \dots
y_m$ and the same error-correcting code with the same BER
$=\alpha$ was applied to it, then after decompression the
resulting BER will be proportional to $n\alpha$ due to the error
propagation effect. Consequently, a more robust
error-correcting code must be used, which means a larger number
of parity bits and increases the decoding time.
This, in turn, can reduce data compression efficiency and
increase encoding and decoding times.

\subsection{The main tool of the proposed method: two-faced processes}

In the main part of the article, we will consider information
sources that generate letters from the alphabet $\{0,1\}$ and,
in particular, a Bernoulli source $B(p)$ that generates letters
with probabilities $P(0) = p, P(1)= q \, (= 1-p), \,\, p \ge 1/2$.

The proposed code will be based on so-called two-faced
processes, which are Markov chains of order (or memory) $s, s
\ge 1$. This means that for the generated sequence $\dots x_{-1}
x_0 x_1 \dots$, the probability $P(x_{i+1})$ depends on the $s$
preceding symbols, i.e., the probability $P(x_{i+1} = \alpha \mid
\dots x_{-1} x_0 x_1 \dots x_i)$ is equal to $P(x_{i+1} = \alpha \mid
x_{i-s+1} \dots x_i)$, where $\alpha \in \{0,1\}$.
We will describe those chains by the following transition
probability matrices:
\begin{equation}\label{ch1}
s=1: \quad
\begin{array}{c|c|c|}
  \text{probability of zero} & p & q \\ 
  \hline 
  \text{after} & 0 & 1 \\
\end{array} \, .
\end{equation}
This means that $Pr\{ x_{i+1} = 0 \mid x_i= 0\} = p$, $Pr\{ x_{i+1} =
0 \mid x_i= 1\} = q$ and, of course,
$Pr\{ x_{i+1} = 1 \mid x_i= 0\} = q$, $Pr\{ x_{i+1} = 1 \mid x_i= 1\} = p$.
It is obvious that $Pr\{ x_{i+1} = 1 \mid \dots\} = 1 - Pr\{ x_{i+1} = 0 \mid \dots\}$, and these probabilities are not contained in the matrices.

\begin{equation}\label{ch2}
\text{For } s=2: \quad
\begin{array}{c|c|c|c|c|}
  \text{probability of zero} & p & q & q & p \\ 
  \hline 
  \text{after} & 00 & 01 & 10 & 11 \\
\end{array} \, .
\end{equation}
This means that $Pr\{ x_{i+1} = 0 \mid x_{i-1}x_i= 00\} = p$, etc. 

The ``typical'' sequences for (\ref{ch1}) and (\ref{ch2}) for
large $p$ (say, $0.9$) can be as follows:
\begin{equation}\label{tt1}
x_1x_2x_3x_4x_5x_6x_7x_8x_9x_{10} \, = \,
0\,\, 0\,\, 0\,\, 0\,\, 1\,\, 1\,\, 1\,\, 1\,\, 1\,\, 1 \,\, ,
\end{equation}
\begin{equation}\label{tt2}
x_1x_2x_3x_4x_5x_6x_7x_8x_9x_{10} \, = \,
0\,\, 0\,\, 0\,\, 0\,\, 1\,\, 1\,\, 0\,\, 1\,\, 1\,\, 0 \, .
\end{equation}
(In both cases, a rare event, i.e., with probability $q$, occurred after $x_4$.)

\section{Intuitive description of the ideas underlying the proposed method}

In this article, we will consider transmission over two binary
channels that are very popular in information theory: the binary
erasure channel (BEC) and the binary symmetric channel (BSC).
For the first channel, any transmitted letter from $\{0,1\}$ is
lost, i.e., the receiver receives a special symbol $*$ (instead
of the transmitted 0 or 1) with a certain probability $\pi$. For
BSC, any letter from the sender can be replaced by another with
a certain probability $\pi$, see \cite{co}.

Now let us assume that a binary sequence is transmitted via BEC
with probability $\pi$, that is, any letter is converted to $*$
during transmission with probability $\pi$.

Let us consider three cases:

i) The transmitted sequence is generated by a Bernoulli source
$B(p), p>1/2$. Obviously, the best strategy is to replace $*$
with $0$, since the probability of $0$ is equal to $p>1/2$. Thus,
$Pr(\text{error}) = q$. In the future, we will use the same strategy
for all cases, that is, we will choose the letter with the
higher posterior probability and, so, $Pr(\text{error})$ will be equal to
the smaller probability.

ii) Consider the case where the sequence generated by a
two-faced process is transmitted over the BEC. Suppose that the
memory of the process is equal to $1$, and the transmitted
sequence (\ref{tt1}) is received as
\begin{equation}\label{t1*}
0\,\, 0\,\, *\,\, 0\,\, 1\,\, 1\,\, 1\,\, *\,\, 1\,\, 1 \, .
\end{equation}
(That is, the third and eighth letters are erased.) Since $x_2=0$ and $x_4=0$, a direct calculation using (\ref{ch1}) and Bayes' rule gives
\[
P(x_3=0 \mid (\ref{t1*})) = $$ $$\frac{P(x_3=0 \mid x_2=0) P(x_4=0 \mid x_3=0)}{P(x_3=0 \mid x_2=0) P(x_4=0 \mid x_3=0) + P(x_3=1 \mid x_2=0) P(x_4=0 \mid x_3=1)} $$ $$= \frac{p^2}{p^2 + q^2}.
\]
Thus, in this case, $Pr(\text{error}) = P(x_3=1 \mid (\ref{t1*})) = q^2/(p^2 + q^2)$. The same is true for the eighth letter.

iii) Now suppose that the memory of the two-faced process is $2$ and
sequence (\ref{tt2}) was transmitted via the same BEC and the
following was received:
\begin{equation}\label{t2*}
0\,\, 0\,\, *\,\, 0\,\, 1\,\, 1\,\, 0\,\, *\,\, 1\,\, 0 \, .
\end{equation}
A direct calculation of conditional probabilities using (\ref{ch2}) shows that 
$P(x_3=0 \mid (\ref{t2*})) = p^3/(p^3 + q^3)$. Thus, in this case, $Pr(\text{error}) = q^3/(p^3 + q^3)$.
The same is true for the eighth letter.

So, from these simple examples, we see that the probability of
error is $q$ for a Bernoulli process and $q^2/(p^2 + q^2)$,
$q^3/(p^3 + q^3)$ for two-faced processes with memory one and two, respectively. Thus, in this
example, we see that increasing the memory of the process
reduces the decoding error (in these examples, we do not
consider all possible channel error positions, but later we will
consider general methods and evaluate their performance).
Thus, in a sense, increasing the memory corrects most of the
channel errors.

The second ingredient of the proposed error correction method is
the availability of a method for transforming the Bernoulli
process (and any other random process) into a two-faced process
with the necessary memory.

In short, the proposed coding scheme is a two-step procedure: first,
the transmitted sequence is converted into a two-faced process
and transmitted over a noisy channel. The resulting data is then
adjusted according to maximum likelihood estimates, and the
resulting sequence is transformed from a two-faced process back
into an estimate of the original process.

For sources without redundancy (such as $B(1/2)$), the stretching code described later in Section VIII uses a three-step procedure: first,
additional symbols (say, zeroes) are added to the original
sequence. Then, the resulting sequence is converted into a
two-faced process, and finally, the sequence received over
the channel is decoded and converted back into an estimate of the
original Bernoulli process (where the decoder knows the positions of the added characters).

\section{Two-faced processes and transformations}

\subsection{Two-faced processes}

Let us define matrices $T_1 = \bigl(\begin{smallmatrix}p&q \\
0&1\end{smallmatrix} \bigr)$, $\overline{T}_1 =
\bigl(\begin{smallmatrix}q&p \\ 0&1\end{smallmatrix} \bigr)$,
$T_2= T_1\parallel \overline{T}_1=
\bigl(\begin{smallmatrix}p&q&q&p \\
00&01&10&11\end{smallmatrix} \bigr)$,
$\overline{T}_2 = \bigl(\begin{smallmatrix}q&p&p&q \\
00&01&10&11\end{smallmatrix} \bigr)$,
$T_3= T_2\parallel \overline{T}_2$, $\dots$, $T_{k+1}=
T_k\parallel \overline{T}_k$, and so on.
For example, 
\[
\overline{T}_3 = \bigl(\begin{smallmatrix}q&p&p&q & p&q&q&p \\
000&001&010&011 &100&101&110&111 \end{smallmatrix} \bigr),
\]
where, as before, $Pr\{ x_{i+1} = 0 \mid x_{i-2}x_{i-1}x_i= 000\} = q$, etc.

More precisely, $\overline{T}_k$ is obtained from $T_k$ by
swapping $p$ and $q$ in the top row of $T_k$, and the upper row of $T_{k+1}$ is a
concatenation of the upper rows of $T_k$ and $\overline{T}_k$,
while the lower row is a sequence of lexicographically ordered
words from $\{0,1\}^{k+1}$.

These processes were proposed by the author in 2004, and a
complete description of them can be found in \cite{rm}. There,
it was proven that for a process with memory $s$, for any $r \le
s$ and any word $g$ of length $r$, $Pr\{x_i \dots x_{i+r-1} = g\} = 2^{-r}$,
whereas the Shannon entropy per symbol $h_\infty$ is equal to $-(p
\log_2 p + q \log_2 q)$, i.e., $h_\infty$ can be very small (for
large $p$). This is why such processes are called two-faced: if
we look at the frequencies of words of length $r \le s$, the
process appears to be uniformly distributed (i.e., ``completely
random''), but if we look at the frequencies of long words (those
longer than $s$), this distribution is very far from uniform.

\subsection{Two-faced transformations}
First, we define two maps $\phi_{l,w} : \{0,1\}^n \to \{0,1\}^n$ and $\psi_{l,w} : \{0,1\}^n \to \{0,1\}^n$,
where integers $n,l$ ($l \le n$) and a word $w \in \{0,1\}^l$ are
parameters.
Let there be given words $w = w_1 \dots w_l \in \{0,1\}^l$ and
$x_1 \dots x_n \in \{0,1\}^n$.

The transformation $\phi_{l,w} : x \to v$ is defined by the initial conditions $v_j = w_{l+j}$ for $j = -l+1, \dots, 0$, and
\[
v_{i} =
  \begin{cases} x_{i} \quad 
\text{if } \sum_{j=i-l}^{i-1} v_j \text{ is even,}
\\
   \overline{x}_{i} \quad 
\text{if } \sum_{j=i-l}^{i-1} v_j \text{ is odd,}
\end{cases}
\]
for $i=1, \dots, n$ (here and below $\overline{0} = 1$, $\overline{1}=0$).

It will be convenient to use the following notations: $a\oplus
b = (a+b) \bmod 2$ and
$\bigoplus_{i=m}^n x_i = x_m \oplus \dots \oplus x_n$.
Thus, the transformation $\phi_{l,w} : x \to v$ can be
rewritten as follows:
\begin{equation}\label{fi}
v_i = x_i \oplus
\bigoplus_{j=i-l\text{\deleted[id=IB]{\(+1\)}}}^{i-1} v_j
\end{equation}
for $i=1, \dots, n$.
For an $n$-bit word $v$ (with $v_{-l+1}\dots v_0 = w_1\dots w_l$), the inverse transformation $\psi_{l,w} : v \to u$ is defined by
\begin{equation}\label{psi} 
u_{i+1} = v_{i+1} \oplus \bigoplus_{j=i-l+1}^i v_j
\end{equation}
for $i=0, \dots, n-1$.
For example, for the sequence $x=0000100000$, $\phi_{1,0}(x)$
equals $0000111111$,
$\phi_{2,00}(x)$ equals $0000110110$, $\psi_{1,0}(0000111111) =
0000100000$, and
$\psi_{2,00}(0000110110) = 0000100000$.

Note that $\psi_{l,w}(\phi_{l,w}(x)) = x$.

The following theorem from \cite{rm} holds for the
transformation $\phi_{l,w}(x)$:
\begin{theorem}
Let $l > 0$ be an integer, $w$ be a uniformly distributed binary
word of $l$ letters, and let $x=x_1\dots x_n$ be generated randomly
according to a Bernoulli process with parameters $p \in (0,1)$.
Then the sequence $\phi_{l,w}(x)$ obeys the distribution 
of a two-faced process $T_l$ of order $l$ (with matrix $T_l$) and
probability $p$.
\end{theorem}

{\it Comment.} In this paper, we will focus on the Bernoulli
process, but a similar theorem holds not only for the Bernoulli
process but also for any stochastic one (see \cite{rm}, Theorem 4).

\section{Code structure}

Let us describe the encoders for BEC and BSC.
In all cases, we will consider the following scheme. There is an
integer $l$ and a word $w \in \{0,1\}^l$, formed according to a
uniform distribution, where this word $w$ and integer $l$ are
known to the sender and receiver.
In addition, there is a word $x=x_1 \dots x_n$ generated by a
Bernoulli source $B(p)$, and this word must be transmitted over
a noisy channel. (Of course, $x$ is unknown to the receiver.)

The case where $l=n$ is of particular interest, since in this
case the codewords $v_1 \ldots v_n$ are uniformly distributed.
(Recall that $w$ is also uniformly distributed. In practice, it
may be difficult to store the same value of $w$ in both the
encoder and the decoder; however, the same pseudorandom number
generator can be used for both.)

Let us denote $\phi_{l,w}(x) = v_1 \dots v_n$. 
Define the word $v= w_1 \dots w_l v_1 \dots v_n$, that is, $v$ is a
word of $(n+l)$ letters, and $v_{-l+1} \dots v_0 = w_1 \dots w_l$.
The word $v_1 \dots v_n$ is transmitted over a noisy channel, and
the receiver receives a distorted word $v^*_1, \dots, v^*_n$.
Note that in the case of BEC, $v^*_i \in \{0,1,*\}$.      
It is convenient to define $v^*_{-l+1} \dots v^*_0$ as
$w_1 \dots w_l$. (This is possible because the word $w$ is known
to the receiver.) Thus, $v^*_{-l+1} \dots v^*_0 = v_{-l+1} \dots v_0$.
  
Decoding is a two-step procedure: first, the sequence $v^* =
v^*_1 v^*_2 \dots v^*_n$ is processed into the sequence $v' = v'_1
v'_2 \dots v'_n$ in the hope of eliminating some errors (but, at
the same time, possibly adding new ones). Then the mapping
$\psi_{l,w}(v')$ is applied, and the resulting sequence $x' =
x'_1 \dots x'_n$ is the result of decoding.
Here, the transformation $v^*_1 v^*_2 \dots v^*_n \to v'_1
v'_2 \dots v'_n$ plays a key role and will be described in the next section.

Another important quantity is the bit error rate (BER), defined
as the average frequency of symbols for which $x'_i \ne x_i$, $i =
1, \dots, n$. We denote it as $BER(Code)$, where $Code$ is a
specific coding method.

\section{Maximum likelihood decoding for BSC and BEC}

\subsubsection{Decoding scheme}

First, let us give a definition. Given the known initial state $v_{-l+1} \dots v_0$ and the received sequence $v^*_1 \dots v^*_n$, the maximum likelihood (MAP) decoder determines the estimate $v'_1 \dots v'_n$ as follows:
\begin{equation}\label{ml} 
v'_1 \dots v'_n = \arg \max_{v_1 \dots v_n \in \{0,1\}^n} P(v_1 \dots v_n \mid v_{-l+1} \dots v_0, \, v^*_1 \dots v^*_n) \,.
\end{equation}
Since the channel is memoryless and $P(v^*_1 \dots v^*_n \mid v_{-l+1} \dots v_0)$ does not depend on $v_1 \dots v_n$, by Bayes' rule and the Markov property of order $l$, maximizing (\ref{ml}) is equivalent to
\begin{equation}\label{ml_prod}
v'_1 \dots v'_n = \arg \max_{v_1 \dots v_n \in \{0,1\}^n} \prod_{i=1}^{n} P(v^*_i \mid v_i) P(v_i \mid v_{i-l} \dots v_{i-1}) \,,
\end{equation}
where $P(v^*_i \mid v_i)$ is the channel transition probability (for BEC, $P(* \mid v_i) = \pi$ and $P(v_i \mid v_i) = 1-\pi$; for BSC, $P(1-v_i \mid v_i) = \pi$ and $P(v_i \mid v_i) = 1-\pi$). If a decoder is designed for a Markov chain of memory $l$, we denote it as $Dec_l$.

\subsubsection{Example: the case where $l=1$}

Let $l=1$, and let us describe the dynamic programming algorithm for computing $v'_1 \dots v'_n$ in (\ref{ml_prod}).

To describe the decoder $Dec_1$, for each step $s = 1, \ldots, n$ and each state $a \in \{0,1\}$ (representing the value of $v_{s-1}$), we define the maximum suffix probability $\mu^a_{n,s}$ and the optimal next symbol $V^a_s \in \{0,1\}$. For $s = n$ and $a \in \{0,1\}$, let
\begin{align}
\mu^a_{n,n} &= \max_{z \in \{0,1\}} \Big\{ P(v^*_n \mid v_n = z) P(v_n = z \mid v_{n-1} = a) \Big\}, \nonumber \\
V^a_n &= \arg \max_{z \in \{0,1\}} \Big\{ P(v^*_n \mid v_n = z) P(v_n = z \mid v_{n-1} = a) \Big\}. \label{mu1}
\end{align}
For $s = n-1, n-2, \ldots, 1$ and $a \in \{0,1\}$, we define recursively
\begin{align}
\mu^a_{n,s} &= \max_{z \in \{0,1\}} \Big\{ P(v^*_s \mid v_s = z) P(v_s = z \mid v_{s-1} = a) \, \mu^z_{n,s+1} \Big\}, \nonumber \\
V^a_s &= \arg \max_{z \in \{0,1\}} \Big\{ P(v^*_s \mid v_s = z) P(v_s = z \mid v_{s-1} = a) \, \mu^z_{n,s+1} \Big\}. \label{mus}
\end{align}

The proposed algorithm $Dec_1$ works in two passes:
\begin{enumerate}
    \item \emph{Backward pass:} It computes $\mu^0_{n,s}, \mu^1_{n,s}, V^0_s, V^1_s$ step by step for $s = n, n-1, \ldots, 1$.
    \item \emph{Forward pass:} Using the known initial symbol $v_0$, it reconstructs the sequence $v'_1 \dots v'_n$ iteratively by setting
    \begin{equation}\label{backtrack1}
    v'_1 = V^{v_0}_1, \qquad v'_s = V^{v'_{s-1}}_s \quad \text{for } s = 2, \ldots, n.
    \end{equation}
\end{enumerate}

The following statement shows that the resulting sequence $v'_1 \dots v'_n$ is an optimal solution of (\ref{ml}) for $l=1$.

\begin{claim}
Let $v = v_1 \ldots v_n$ be the input sequence, and $v_1^* \ldots v_n^*$ be the output sequence of a memoryless noisy channel, where $v$ is generated by a Markov chain (\ref{ch1}) with known initial state $v_0$. Then the output sequence $v'_1 \dots v'_n$ of the decoder $Dec_1$ defined in (\ref{backtrack1}) is a solution of (\ref{ml}).
\end{claim}

\begin{proof}
For any $s \in \{1, \ldots, n\}$ and $a \in \{0,1\}$, let us show by backward induction on $s$ that
\begin{equation}\label{ind_hyp}
\mu^a_{n,s} = \max_{b_s \dots b_n \in \{0,1\}^{n-s+1}} \prod_{i=s}^{n} P(v^*_i \mid v_i = b_i) P(v_i = b_i \mid v_{i-1} = b_{i-1}),
\end{equation}
where $b_{s-1} = a$, and that the sequence $b'_s \dots b'_n$ constructed via $b'_s = V^a_s$ and $b'_i = V^{b'_{i-1}}_i$ ($i = s+1, \ldots, n$) achieves this maximum.

For the base case $s = n$, (\ref{ind_hyp}) holds directly by definition (\ref{mu1}). Suppose that the claim holds for $s+1$ ($1 \le s \le n-1$). Then for step $s$ and any $a \in \{0,1\}$ (with $b_{s-1} = a$), we have
\begin{align*}
&\max_{b_s \dots b_n \in \{0,1\}^{n-s+1}} \prod_{i=s}^{n} P(v^*_i \mid v_i = b_i) P(v_i = b_i \mid v_{i-1} = b_{i-1})  =   \max_{b_s \in \{0,1\}}
 \\
& \left\{ P(v^*_s \mid v_s = b_s) P(v_s = b_s \mid v_{s-1} = a)
 \max_{b_{s+1} \dots b_n \in \{0,1\}^{n-s}} \prod_{i=s+1}^{n} P(v^*_i \mid v_i = b_i) P(v_i = b_i \mid v_{i-1} = b_{i-1}) \right\} \\
&= \max_{b_s \in \{0,1\}} \Big\{ P(v^*_s \mid v_s = b_s) P(v_s = b_s \mid v_{s-1} = a) \, \mu^{b_s}_{n,s+1} \Big\} = \mu^a_{n,s},
\end{align*}
where the second equality follows from the induction hypothesis, and the last equality follows from (\ref{mus}). Moreover, by (\ref{mus}), the outer maximum is achieved at $b'_s = V^a_s$, and by the induction hypothesis, the inner maximum is achieved by continuing with $b'_i = V^{b'_{i-1}}_i$ for $i = s+1, \ldots, n$.

Setting $s = 1$ and $a = v_0$ in (\ref{ind_hyp}), we conclude that $v'_1 \dots v'_n$ given by (\ref{backtrack1}) maximizes (\ref{ml_prod}), and therefore is a solution of (\ref{ml}).
\end{proof}

\subsubsection{General case: $l > 1$}

We now generalize the algorithm above to any memory order $l \ge 1$. For a binary word $u = u_1 u_2 \dots u_l \in \{0,1\}^l$ and a symbol $z \in \{0,1\}$, let $\sigma(u, z) = u_2 \dots u_l z \in \{0,1\}^l$ denote the shifted window of length $l$ (for $l=1$, $\sigma(u, z) = z$).

For each step $s = 1, \ldots, n$ and each state $u \in \{0,1\}^l$ (representing the context $v_{s-l} \dots v_{s-1}$), we define the real numbers $\mu^u_{n,s}$ and the optimal next symbols $V^u_s \in \{0,1\}$. For $s = n$ and $u \in \{0,1\}^l$, let
\begin{align}
\mu^u_{n,n} &= \max_{z \in \{0,1\}} \Big\{ P(v^*_n \mid v_n = z) P(v_n = z \mid v_{n-l} \dots v_{n-1} = u) \Big\}, \nonumber \\
V^u_n &= \arg \max_{z \in \{0,1\}} \Big\{ P(v^*_n \mid v_n = z) P(v_n = z \mid v_{n-l} \dots v_{n-1} = u) \Big\}. \label{mu1l}
\end{align}
For $s = n-1, n-2, \ldots, 1$ and $u \in \{0,1\}^l$, we define
\begin{align}
\mu^u_{n,s} &= \max_{z \in \{0,1\}} \Big\{ P(v^*_s \mid v_s = z) P(v_s = z \mid v_{s-l} \dots v_{s-1} = u) \, \mu^{\sigma(u, z)}_{n,s+1} \Big\}, \nonumber \\
V^u_s &= \arg \max_{z \in \{0,1\}} \Big\{ P(v^*_s \mid v_s = z) P(v_s = z \mid v_{s-l} \dots v_{s-1} = u) \, \mu^{\sigma(u, z)}_{n,s+1} \Big\}. \label{musl}
\end{align}

The decoder $Dec_l$ operates as follows:
\begin{enumerate}
    \item \emph{Backward pass:} It computes $\mu^u_{n,s}$ and $V^u_s$ for all $u \in \{0,1\}^l$ step by step for $s = n, n-1, \ldots, 1$.
    \item \emph{Forward pass:} Using the known initial state $v_{-l+1} \dots v_0$ (setting $v'_j = v_j$ for $j = -l+1, \ldots, 0$), it reconstructs $v'_1 \dots v'_n$ iteratively by
    \begin{equation}\label{backtrackl}
    v'_s = V^{v'_{s-l} \dots v'_{s-1}}_s \quad \text{for } s = 1, \ldots, n.
    \end{equation}
\end{enumerate}

\begin{theorem}
Let $v = v_1 \ldots v_n$ be the input sequence, and $v_1^* \ldots v_n^*$ be the output sequence of a memoryless noisy channel, where $v$ is generated by a Markov chain with memory $l \ge 1$ and known initial state $v_{-l+1} \dots v_0$. Then:
\begin{enumerate}
    \item[i)] The output sequence $v'_1 \dots v'_n$ of the decoder $Dec_l$ given by (\ref{backtrackl}) is a solution of (\ref{ml}).
    \item[ii)] The runtime of $Dec_l$ is $O(2^l n)$.
\end{enumerate}
\end{theorem}

\begin{proof}
Statement i) follows by repeating the backward induction proof of Claim 1 with the state $a \in \{0,1\}$ replaced by the $l$-bit context $u \in \{0,1\}^l$, the state transition $b_s$ replaced by the window shift $\sigma(u, b_s)$, and equations (\ref{mu1l})--(\ref{backtrackl}) used instead of (\ref{mu1})--(\ref{backtrack1}).

For statement ii), at each of the $n$ steps $s = n, \ldots, 1$, the backward pass evaluates (\ref{musl}) for all $2^l$ words $u \in \{0,1\}^l$, performing $O(1)$ arithmetic operations per word. The forward pass (\ref{backtrackl}) takes $O(n)$ operations. Hence, the total runtime is $O(2^l n)$.
\end{proof}

\clearpage
\section{Experiments}

\begin{figure}[p]
  \centering
\includegraphics[width=\linewidth,height=0.88\textheight,keepaspectratio]{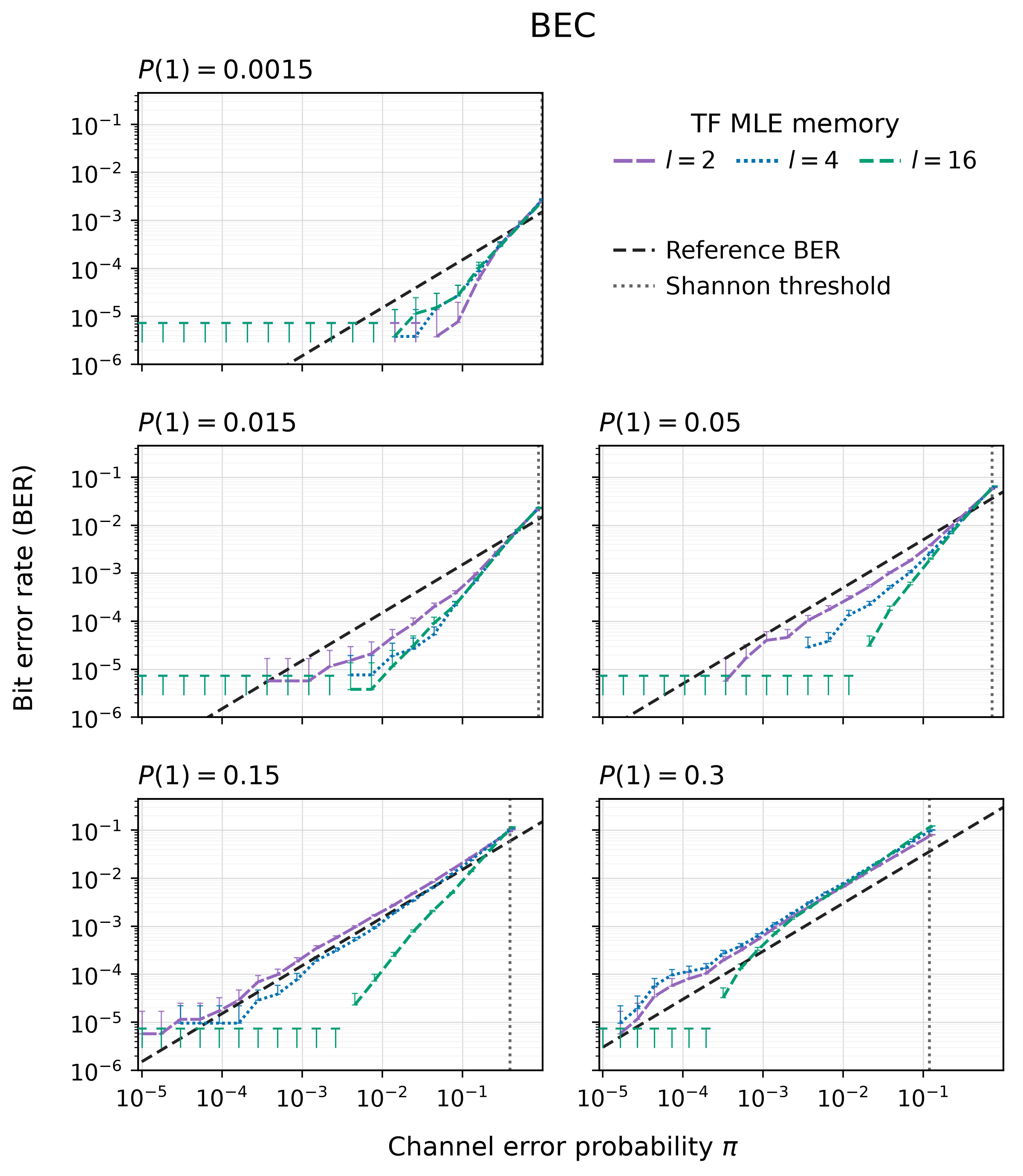}
\caption[TF MLE on the BEC]{\protect\added[id=IB]{BER for TF MLE
on the BEC.}}
  \label{fig:tf-mle-bec}
\end{figure}

\begin{figure}[p]
  \centering
\includegraphics[width=\linewidth,height=0.88\textheight,keepaspectratio]{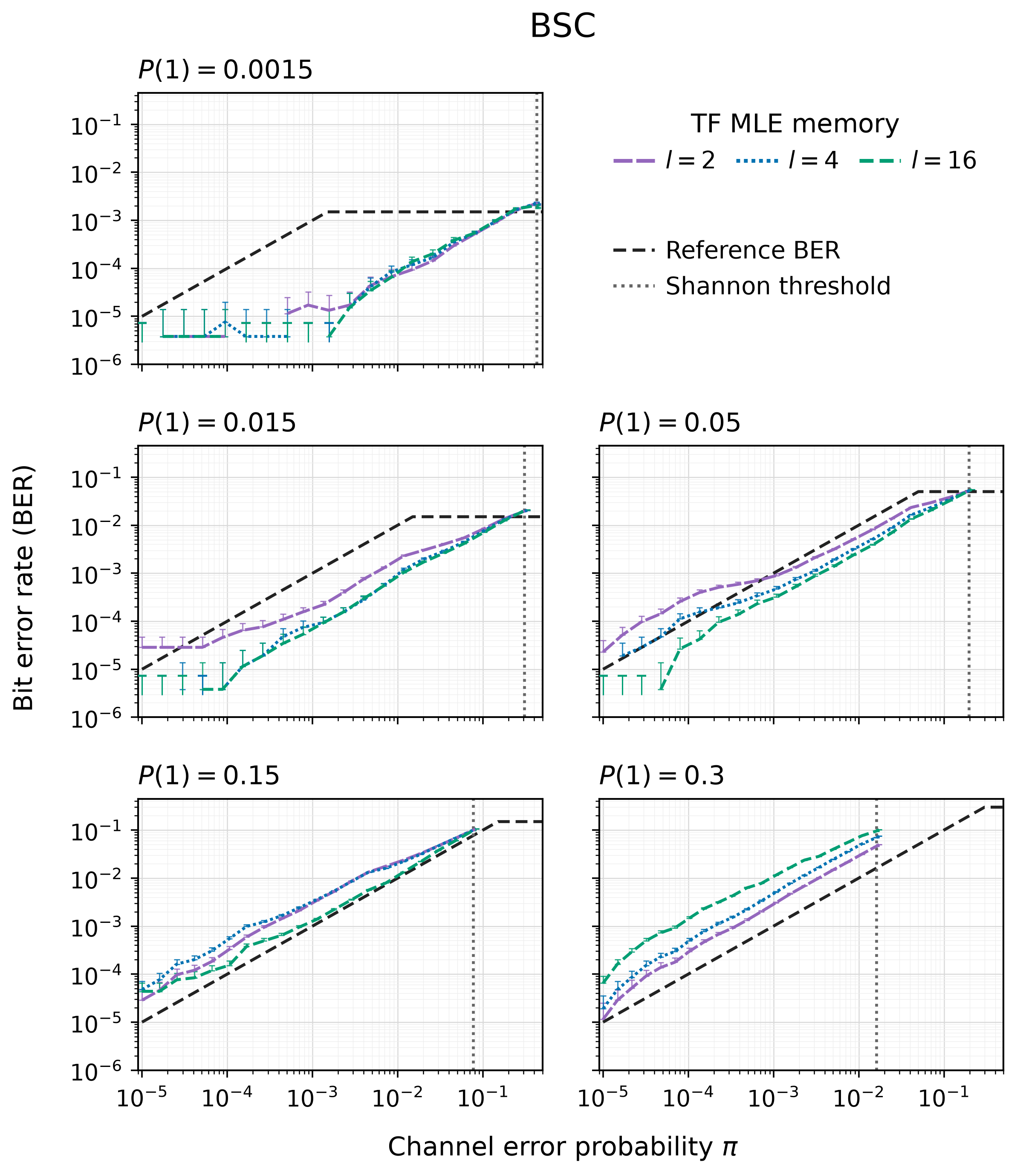}
\caption[TF MLE on the BSC]{\protect\added[id=IB]{BER for TF MLE
on the BSC.}}
  \label{fig:tf-mle-bsc}
\end{figure}

To compare codes practically, we conducted several experiments
of programmatic implementation of described codes.
We measured bit error rate (BER) for erasure (BEC) and
symmetrical (BSC) channels.
For BEC, channel error probability $\pi$ is probability that
symbol will be erased, and for
BSC $\pi$ is probability that symbol will be flipped.
Also, in this section we assume $p=P(1)$. 
Figures~\ref{fig:tf-mle-bec} and~\ref{fig:tf-mle-bsc} show the
measured bit error rate (BER) for
$p\in\{0.0015,0.015,0.05,0.15,0.3\}$
and memory orders $l\in\{2,4,16\}$. 
For each source probability and channel, we tested all possible
values of the channel error probability $\pi$.
For each test, we encoded and decoded $k=256$ blocks of $n=2048$
bits and checked correctness.
Error bars show the upper endpoints of 95\% Wilson
intervals~\cite{wilson}, including when no errors were
observed.

Plots are based on data which is published at
\href{https://github.com/zpix1/twofaced-decoding-methods-data}{\nolinkurl{github.com/zpix1/twofaced-decoding-methods-data}}.

For the BEC, the dashed reference line is $p\pi$, the uncoded
BER when the receiver replaces erased bits with zero.

For the BSC, the dashed reference line gives the optimal uncoded
BER $\min(\pi,p)$ for $p\le 1/2$ and $0\le\pi\le 1/2$. The
uncoded receiver keeps the received bit when $\pi\le p$ and
always outputs zero when $\pi>p$.

For each $p$, Tables~\ref{tab:bec-gain} and~\ref{tab:bsc-gain}
show intervals where code was better than reference.

\noindent\begin{minipage}[t]{0.48\columnwidth}
  \centering
  \footnotesize
  \setlength{\tabcolsep}{4pt}
\captionof{table}{\protect {BEC improvement range
(relative to $p\pi$)}}
  \label{tab:bec-gain}
  \begin{tabular}{c c c}
  \hline
$p$ &  Gain range in $\pi$ &
$l$ \\
  \hline
$0.0015$ & $\le 0.2933$ &
 $2,4,16$ \\
$0.015$& $\le 0.2675$ &
 $4,16$ \\
 $0.05$ &  $\le 0.2201$&
 $2,4,16$\\
 $0.15$ &  $\le 0.1282$ &
 $16$ \\
 $0.3$ &  $\le 5.195\times10^{-4}$ &
$16$ \\
  \hline
  \end{tabular}

\end{minipage}\hfill
\begin{minipage}[t]{0.48\columnwidth}
  \centering
  \footnotesize
  \setlength{\tabcolsep}{4pt}
\captionof{table}{\protect BSC improvement range
(relative to $\min(\pi,p)$)}
  \label{tab:bsc-gain}
  \begin{tabular}{c c c}
  \hline
 $p$ &  Gain range in $\pi$ &
 $l$ \\
  \hline
 $0.0015$ & $\le 0.1385$ &
 $2,4,16$ \\
 $0.015$ & $\le 0.1773$ &
 $4,16$ \\
 $0.05$ &  $\le 0.1164$ &
$16$ \\
  \hline
  \end{tabular}
\end{minipage}

\clearpage
\section{Expanding the Scope of Application}

\subsubsection{Extension of the Bernoulli distribution $B(p)$ for any value of $p$}
The experimental results in Section VII show that the described code does not reduce
the error rate for all pairs of source and channel parameters.
For example, if $p=1/2$, the proposed code---like any other---cannot
be applied without increasing the length of the input sequence
(typically by adding parity bits).

Obviously, the sender cannot change the channel error value. The
only option is to encode the output sequence in such a way as to
change the value of the parameter $p$. (Of course, it must be
possible to decode the encoded sequence back.)

So, we describe a code designed to stretch
the original sequence satisfying the condition $B(1/2)$ in such
a way that the encoded sequences are close to the sequences
generated by $B(p), p \ne 1/2$, with the required $p$.

Let us consider the transformation of sequences from $B(1/2)$
to $B(p)$, where $p \in (0,1/2)$.
First, let us describe the coding scheme. Let the initial
sequence $x = x_1 \ldots$ be generated by a Bernoulli source
$B(1/2)$. The sequence $x$ is divided into sub-sequences $y_1 =
x_1 \ldots x_m$, $y_2 = x_{m+1} \ldots x_{2m}$, $\dots$, where
the integer $m$ is a parameter of the method. Then we
define the set of $k$-letter binary words with $s$ ones as
$C(k,s)$ and determine the minimum $k$ for which $2^m \le |C(k,
\lfloor kp\rfloor )|$. Then the encoder $\rho$ is a mapping from
$\{0,1\}^m \to C(k, \lfloor kp\rfloor)$, whereas the decoder
$\rho^{-1}$ is such a mapping from the set
$\rho(\{0,1\}^{\text{\replaced[id=IB]{$m$}{$k$}}} ) $ into $
\{0,1\}^m$ that $\rho^{-1} (\rho (u) )= u$ for any $u \in
\{0,1\}^m$.

The decoding and encoding problems are well known in information
theory, where they are often referred to as enumeration
problems. For both problems, there exists an algorithm with a
computational time and memory complexity of $O(m\log m)$ (see
\cite{ba} or the general enumeration method in \cite{ry2}).

\subsubsection{Extension to any stochastic processes}
Let us briefly and informally discuss how the results obtained
can be applied to general stochastic processes. As we mentioned
above, the two-faced transformation can be applied to any
stochastic process, so the encoder can be used as described
above. The described maximum likelihood decoding process can be
applied directly if the decoder uses the conditional
probabilities of the process.

\section{Conclusion}

The error-correcting code proposed here performs encoding and decoding at a constant rate and requires a memory capacity of approximately $2^8 \text{--} 2^{16}$ cells, which is quite acceptable for modern processors. Given this, it can be said that its use requires virtually no additional resource overhead. Thus, this code can be used when transmitting redundant data over low-noise channels to reduce the bit error rate (BER) by a factor of several, while employing relatively simple encoders and decoders. This simplicity allows the code to be used in combination with other joint source-channel coding (JSCC) methods, though this may require further research.

\end{document}